\documentclass[reprint,prl,twocolumn,superscriptaddress,showpacs,nofootinbib,floatfix,preprintnumbers]{revtex4-1}
\usepackage{multirow}
\usepackage{diagbox}
\usepackage{graphicx,bm, color, amssymb, amsmath,subfigure}

\newcommand{\be}{\begin{eqnarray}}
\newcommand{\ee}{\end{eqnarray}}
\begin{document}
\preprint{}
\title{Limiting the  Effective Mass and New Physics Parameters from  $0\nu \beta \beta$}

\newcommand{\iiserm}{\affiliation{Indian Institute of Science Education and Research Mohali, Knowledge City, Sector 81, SAS Nagar, Manauli 140306, India}}

\newcommand{\iop}{\affiliation{Institute of Physics, Sachivalaya Marg, 
Bhubaneswar 751005, India}}

\newcommand{\ippp}{\affiliation{
Institute for Particle Physics Phenomenology (IPPP), Durham University,  South Road, Durham DH1 3LE, UK}}

\author{Ram Lal Awasthi}\iiserm
\author{Arnab Dasgupta}\iop
\author{Manimala Mitra}\iiserm\ippp

\begin{abstract}
In the light of the recent result from KamLAND-Zen (KLZ) and GERDA Phase-II, we update the  bounds on the effective mass and the new physics parameters, relevant for neutrinoless double beta decay
($0\nu \beta \beta$). In addition to the light Majorana  neutrino exchange, we  analyse  beyond standard model contributions that arise in   Left-Right symmetry  and R-Parity violating supersymmetry. 
The improved limit from KLZ constrains  the effective mass of 
 light neutrino exchange down to  sub-eV mass regime  0.06 eV. Using the correlation
 between the $^{136}\rm{Xe}$ and $^{76}\rm{Ge}$ half-lives,  we show that the  KLZ limit  individually rules out
the positive claim of observation of $0\nu\beta\beta$ for all nuclear matrix element compilation. For the Left-Right symmetry and R-parity violating supersymmetry,  the KLZ
bound implies a factor of 2 improvement of the effective mass and the new physics  parameters. The future  
ton scale experiments such as, nEXO will  further constrain these models, in particular, 
will rule out standard as well as Type-II dominating LRSM inverted hierarchy scenario.

\end{abstract}
\maketitle
\textit {\textbf {Introduction}}-- The experimental observations of neutrino mass and mixing have opened a new window to physics beyond the standard model (SM).
So far, the solar and atmospheric mass square differences ($\Delta m^2_{21}$ and  $\Delta m^2_{31}$),  the three oscillation angles
$\theta_{12}$, $\theta_{23}$ and $\theta_{13}$ have been measured to a moderate degree of precision \cite{Gonzalez-Garcia:2015qrr}. The  remaining open 
questions in the leptonic sector,  that still need to be  answered are:  the neutrino mass hierarchy and the lightest neutrino mass scale,  the CP-violating phases  and  the
fundamental nature  of SM neutrinos - if  they are Dirac or Majorana particle.  The Majorana mass of the light neutrinos violates
lepton number conservation, and hence, 
this can be determined by observing lepton number violating (LNV) signature 
in  neutrinoless double beta decay ($0\nu \beta \beta$)  $(A,Z) \to (A,Z-2)+2e^{-}$ \cite{0nu2beta-old}.

In a concrete model that generates viable neutrino mass and mixing,  beyond standard model (BSM) states carrying LNV  can also 
directly mediate  $0\nu \beta \beta$. These  additional contributions have been  widely discussed in the literature  
 \cite{feinberg}, 
in particular, for Left-Right symmetry (LRSM) (\cite{LR} and \cite{Tello:2010,Chakrabortty:2012mh,Barry:2013xxa,Suhonen:1998ck,Dev:2013vxa, Dev:2014xea, Bambhaniya:2015ipg, Awasthi:2015ota, Dell'Oro:2015tia, Ge:2015yqa, Dev:2013oxa}),  for R-parity violating supersymmetry (RPV) 
\cite{msnu2beta, mwex, pionex, hirsch, allanach, vogel, Allanach:2014lca, Allanach:2014nna, Dev:2016uxj}, and for other 
scenarios \cite{choi, Mitra:2011qr}. The BSM states of mass  within a few tens of TeV  can significantly contribute to 
$0\nu \beta \beta$ \cite{Dev:2013vxa,Mitra:2011qr, Allanach:2014lca, Allanach:2014nna} and  saturate the present experimental limits, while being in 
accordance with the 
collider \cite{ATLAS:2016lvi, Helo:2015ffa, Aad:2015xaa, Khachatryan:2014dka, Allanach:2014lca, Allanach:2014nna, Deppisch:2015qwa, 
ATLAS:2015gma, ATLAS:2016nij, ATLAS:2016soo, ATLAS:2015jla, Chatrchyan:2013xsw, Khachatryan:2014ura} and cosmological bounds \cite{Ade:2015xua}.

Several  experimental searches have been  carried out till date
to look for the signal in $0\nu \beta \beta$. The  recent bound on the half-life of $^{136}\rm{Xe}$ as reported  by KamLAND-Zen  $T^{0\nu}_{1/2}=1.07\times 10^{26} \,\rm{yrs}$
(90$\%$ C.L) \cite{KamLAND-Zen:2016pfg} provides  almost one order of magnitude improvement  compared to the previous bounds: $T^{0\nu}_{1/2}=1.9\times 10^{25} \,\rm{yrs}$ (KLZ 90$\%$ C.L)
\cite{Gando:2012zm}, 
$T^{0\nu}_{1/2}=1.6\times 10^{25} \,\rm{yrs}$ (EXO-200 90$\%$ C.L) \cite{Albert:2014awa}, $T^{0\nu}_{1/2}=3.4\times 10^{25} \,\rm{yrs}$ (KLZ+EXO 90$\%$ C.L) \cite{Gando:2012zm}. The present limit on the half-life of $^{76}\rm{Ge}$ is $T^{0\nu}_{1/2}=5.2 \times 10^{25} \,\rm{yrs}$ (GERDA Phase-II 90$\%$) \cite{gerda2}. The previous limits are $T^{0\nu}_{1/2}=2.1\times 10^{25} \,\rm{yrs}$ (GERDA 90$\%$) and 
$T^{0\nu}_{1/2}=3.0\times 10^{25} \,\rm{yrs}$ (GERDA+Heidelberg-Moscow+IGEX  90$\%$) \cite{Agostini:2013mzu}. There has been only one claim of
observation of $0\nu \beta \beta$ with the half-life $T^{0\nu}_{1/2}=2.23^{+0.44}_{-0.31}\times 10^{25} \,\rm{yrs}$ (68$\%$ C.L) \cite{Klapdor:2006ff}, which has been significantly constrained by the the measurements from  GERDA and KLZ \cite{gerda2,Dev:2013vxa}. 

In the light of the recent KamLAND-Zen result \cite{KamLAND-Zen:2016pfg} and result from GERDA Phase-II \cite{gerda2},  we re-analyze the different contributions in $0\nu \beta \beta$ that arise  in   LRSM and RPV susy scenarios. We consider both the canonical light neutrino  and  
 BSM exchange mechanisms in $0\nu \beta \beta$, such as  a)  right handed (RH) gauge boson and  right handed  neutrino exchange
in LRSM and b) sbottom and gluino exchange in RPV susy and derive the updated limits on the relevant parameters.  
In addition, we re-check the validity of the positive claim of observation against the  null result of KLZ  and show 
 that assuming the light neutrino exchange as the only mechanism of $0\nu \beta \beta$, 
the recent  KLZ limit  completely  rules out
the positive claim of observation of $0\nu \beta \beta$ for all nuclear matrix elements (NMEs), and so do the new limit from GERDA Phase-II. For LRSM and RPV susy, 
we further explore the  prediction of these theories in future ton-scale experiments, such as, nEXO.


{\textit{\textbf{Left-Right Symmetry:}}} The Left-Right symmetry \cite{LR} is one of the most appealing renormalizable framework, that
 can explain
 light neutrino mass and mixing via a combination of Type-I \cite{type1} and Type-II Seesaw \cite{type2}. The model consists of 
the $SU(2)_L$  doublets 
- $Q_L\equiv (u ~~ d)^{\sf T}_L$ and $\psi_L\equiv (\nu_\ell ~~ \ell)^{\sf T}$, 
$SU(2)_R$ doublets $Q_R\equiv (u ~~ d)^{\sf T}_R$ and $\psi_R\equiv (N_R ~~ l_R)^{\sf T}$. The  Higgs sector of the model consists of 
a bidoublet  $\Phi$ and $SU(2)_{L (R)}$-triplets $\Delta_{L (R)}$. The generic Yukawa Lagrangian of the model is given by
\begin{align}
{\cal L}_Y \  = \ & h_{q}\overline{Q}_{L}\Phi Q_{R}+\tilde{h}_{q}\overline{Q}_{L}\widetilde{\Phi} Q_{R}+
h_{l}\overline{\psi}_{L}\Phi \psi_{R}  + \tilde{h}_{l}\overline{\psi}_{L}\widetilde{\Phi}\psi_{R} \nonumber \\
&
 +f_{L} \psi^C_{L}\Delta_L \psi_{L}+f_{R}\psi^C_{R}\Delta_R \psi_{R} + {\rm H.c.}.
\label{eq:yuk}
\end{align}
In the above,  $C$ denotes  charge conjugation operator and $\widetilde{\Phi}=\tau_2\Phi^*\tau_2$, where $\tau_2$ is  the second Pauli matrix. The above Lagrangian generate the Dirac mass  of the light neutrinos after electroweak symmetry breaking by the bidoublet vacuum expectation value (VEV) $\langle\Phi\rangle={\rm diag}(\kappa, \kappa')$,  $M_D = h_{l}\kappa + \tilde{h}_{l}\kappa'$.  The triplet VEVs of $\langle\Delta^0_{L,R}\rangle$ (denoted as $v_{L,R}$) generate the Majorana mass terms  of light neutrino and heavy neutrino $m_L=f_Lv_L$ and $M_R=f_Rv_R$, respectively. In the seesaw approximation, the light neutrino mass matrix 
\begin{align}
M_\nu \ \simeq \ m_L-M_DM_R^{-1}M_D^{\sf T} \;.
\label{mass}
\end{align} 
In the above, the first and second terms represent  the Type-II (Type-I) seesaw contributions. One of the simplistic possibility is the 
Type-II seesaw dominance in  the light neutrino mass matrix that leads to $M_\nu \sim f_L v_L = v_L M_R/v_R$. This 
occurs as a consequence of   $f_L=f_R$ (or $f_L=f_R^*$), which can be realized as an artifact of  parity (charge conjugation ) 
 symmetry of the Lagrangian.   The other regime of Type-I seesaw dominance
can be realized for vanishingly small triplet vev $v_L  \sim 0$, and the light neutrino mass in this case is 
$
M_\nu \ \simeq -M_DM_R^{-1}M_D^{\sf T} \;.
$\\

{\textit{\textbf{Neutrinoless Double Beta Decay}}}-- In  LRSM several new contributions arise that are mediated by the RH gauge boson, RH neutrino and Higgs triplet  \cite{Tello:2010, Pantis:1996py, review}. The half life $T^{0\nu}_{1/2}$  is given by: 
\begin{eqnarray}
\frac{1}{T_{1/2}^{0\nu}} \ = \ G_{0\nu}g_A^4 \left| \sum_i \mathcal{M}_i  \eta_i \right|^2, 
\label{halft}
\end{eqnarray}
where $G_{0\nu}$ is the phase space factor, $g_A$ is the nucleon  axial-vector coupling constant,  ${\cal M}_{i}$ represents  the NMEs for the different exchange processes, and $\eta_i$ are the corresponding dimensionless particle physics parameters.  Below, we discuss different contributions.

\begin{itemize}
\item
{\bf{Standard light neutrino exchange:}} The light neutrino, if Majorana, can 
mediate the $0\nu \beta \beta$ process. The dimensionless parameter is, 
\be
\eta_\nu & = & \frac{1}{m_e}\sum_i U_{ei}^2 m_i=\frac{m^{\nu}_{ee}}{m_e}  \; , \label{etanu}
\ee
where $m^{\nu}_{ee}$ is the effective mass for light neutrino exchange and $m_e$ is the electron mass.
\item
{\bf{RR contribution:}} The contribution from  $W_R$ and $N_R$ exchange is one of the dominant contribution in $0\nu \beta \beta$, that 
 depends only on the masses of 
the intermediate states and the gauge coupling. For a generic $g_R \neq g_L$, and generic  RH neutrino mass the  
 dimensionless parameter is {(see see \cite{Helo:2010cw} and \cite{Faessler:2014kka} for the validity of this expression)}: 
\be
\eta^R_{N_R} & \sim & m_p \left(\frac{g_R}{g_L}\right)^4\left(\frac{M_{W_L}}{M_{W_R}}\right)^4 \sum_i  \frac{{V_{ei}^*}^2M_i}{|p^2|+M^2_i} \; . \label{etaRR} 
\ee
In the above,  $V_{ei}$ are the elements of the unitary matrix that  diagonalizes the RH neutrino mass matrix $M_R$ with eigenvalues $M_i$,
$|p| \sim 100$ MeV is the typical  momentum exchanged  scale of $0\nu\beta\beta$ with $|p^2|=m_em_p \frac{\mathcal{M}_{N}}{\mathcal{M}_{\nu}}$. Here, $m_e$, $m_p$ are the masses of electron and proton, $\mathcal{M}_{\nu}$ and $\mathcal{M}_N$ are the NME corresponding to the
light and heavy neutrino exchange, respectively.

\item
{\bf{LL contribution:}}
 The $W_L-N-W_L$ mediated LL contribution for generic mass $M_i$, smaller or larger than the momentum exchange scale is 
\be
\eta^L_{N_R}  &  = & m_p \sum_i \frac{S^2_{ei} M_i}{|p^2|+M^2_i} \;. \label{etaRL} 
\ee
where $S_{ei}$ is the element of  the active-sterile mixing matrix.

In addition, few other contributions that depend on the $W_L-W_R$ mediation/mixing can significantly contribute for large contribution to $0\nu \beta \beta$ \cite{Suhonen:1998ck} . They are:
\item
{\bf{$\lambda$ contribution:}}
The light neutrino, $W_L$ and $W_R$  mediated $\lambda$ contribution can be large for large  active-sterile neutrino mixing $T$. The relevant dimensionless parameter is:  
\be
\eta_{\lambda} &=&  \left(\frac{M_{W_L}}{M_{W_R}}\right)^2 \sum_i U_{ei} T^*_{ei} \; , \label{etalam} 
\ee

\item
{\bf{$\eta$ contribution:}}
The light neutrino mediated  $\eta$ contribution depends on the $W_L-W_R$ mixing parameter $\xi$ and can be large for large $\xi$ \cite{Barry:2013xxa, Dev:2014xea}. The dimensionless parameter is 
\be
\eta_{\eta} &=&  \tan \xi \sum_i U_{ei} T^*_{ei} \; , \label{etaeta}
\ee
Note that, for TeV scale RH neutrino, their contribution in $\lambda$ and $\eta$ diagrams are  small and can be ignored.

\item
{\bf{Triplet exchange:}} The corresponding dimensionless parameter for right triplet  exchange is
\be
\eta_{\Delta_R}=\frac{m_p}{G^2_F} \frac{\sum_i V^2_{ei}M_i}{M^4_{W_R} m^2_{\Delta_R}},
\ee
where $M_i$ are the masses of the RH neutrino and $m_{\Delta_R}$ is the mass of the RH doubly charged Higgs triplet. The 
RH triplet of mass comparable or lower than the RH neutrino can give significant contribution in $0\nu\beta\beta$ that together 
with LFV processes can significantly constrain the quasi-degenerate regime  \cite{Bambhaniya:2015ipg}.
The left handed triplet contribution is proportional to the light neutrino masses and therefore  small. Hence, we do not 
consider this into our compilation. 

\begin{table}[!h]
\centering
\begin{tabular}{|c|c|c|c|c|}
\hline
\multicolumn{3}{|c|}{ NME } & \multirow{2}{*}{$|m^{\nu}_{ee}|$}&\multirow{2}{*}{$|m^{\nu}_{ee}|$}  \\ \cline{1-3}
Method & $\mathcal{M}^{0\nu}$ & $\mathcal{M}^{0\nu}$ & & \\ 
& $(^{76}\textrm{Ge})$ &  $(^{136}\textrm{Xe})$ & $(^{76}\textrm{Ge})$ & $(^{136}\textrm{Xe})$\\ \hline
EDF(U)\cite{Rodriguez:2010mn} & 4.6 & 4.2 & 0.20 & 0.06\\ 
ISM(U) \cite{Menendez:2008jp}& 2.81 & 2.19 & 0.33 & 0.12 \\ 
IBM-2 \cite{Barea:2013bz} & 5.42 & 3.33 & 0.17 & 0.08\\ 
pm-QRPA(U)\cite{Suhonen:2010zzc} & 5.18 & 3.16 & 0.18 & 0.08\\ 
SRQRPA-B\cite{Meroni:2012qf} & 5.82 & 3.36 & 0.16 & 0.08\\ 
SRQRPA-B\cite{Meroni:2012qf} & 4.75 & 2.29 & 0.20 & 0.11\\ 
QRPA-B \cite{Simkovic:2013qiy}& 5.57 & 2.46 & 0.17 & 0.11\\ 
QRPA-A\cite{Simkovic:2013qiy} & 5.16 & 2.18 & 0.18 & 0.12 \\ 
SkM-HFB-QRPA \cite{Mustonen:2013zu}& 5.09 & 1.89 & 0.18 & 0.14 \\ 
\hline
\end{tabular}
\caption{The limits on the effective mass $|m^{\nu}_{ee}|$ for light neutrino exchange, that satisfy the KLZ bound $T^{0\nu}_{1/2}=1.07 \times 10^{26} \rm{yrs}$ \cite{KamLAND-Zen:2016pfg} and the limit from GERDA Phase-II $T^{0\nu}_{1/2}= 5.2 \times 10^{25} \rm{yrs}$ \cite{Majorovits:2015vka}.  \label{tab:tab1}}
\end{table}
\end{itemize}

Following the recent  limit of half-life $T^{0\nu}_{1/2}$ from KLZ \cite{KamLAND-Zen:2016pfg}, we show the bound  on the effective mass parameter for standard light neutrino exchange mechanism in  Table~\ref{tab:tab1}. We adopt  the phase space factor  from \cite{Kotila:2012zza}, and  include the NME uncertainties in our compilation 
\cite{Rodriguez:2010mn, Menendez:2008jp, Barea:2013bz, Suhonen:2010zzc, Meroni:2012qf, 
Simkovic:2013qiy, Mustonen:2013zu}.
 Additionally, we also show the limits 
from  GERDA Phase-II \cite{gerda2,Majorovits:2015vka}. The recent KLZ limit constrains the effective mass $|m^{\nu}_{ee}|$ in the sub-eV regime $m^{\nu}_{ee} \leq 0.06-0.14$ eV, almost a factor of 2 
improvement as compared to the previous limit \cite{Dev:2013vxa}.

\begin{figure}[!htb]
\includegraphics[width=8.70cm]{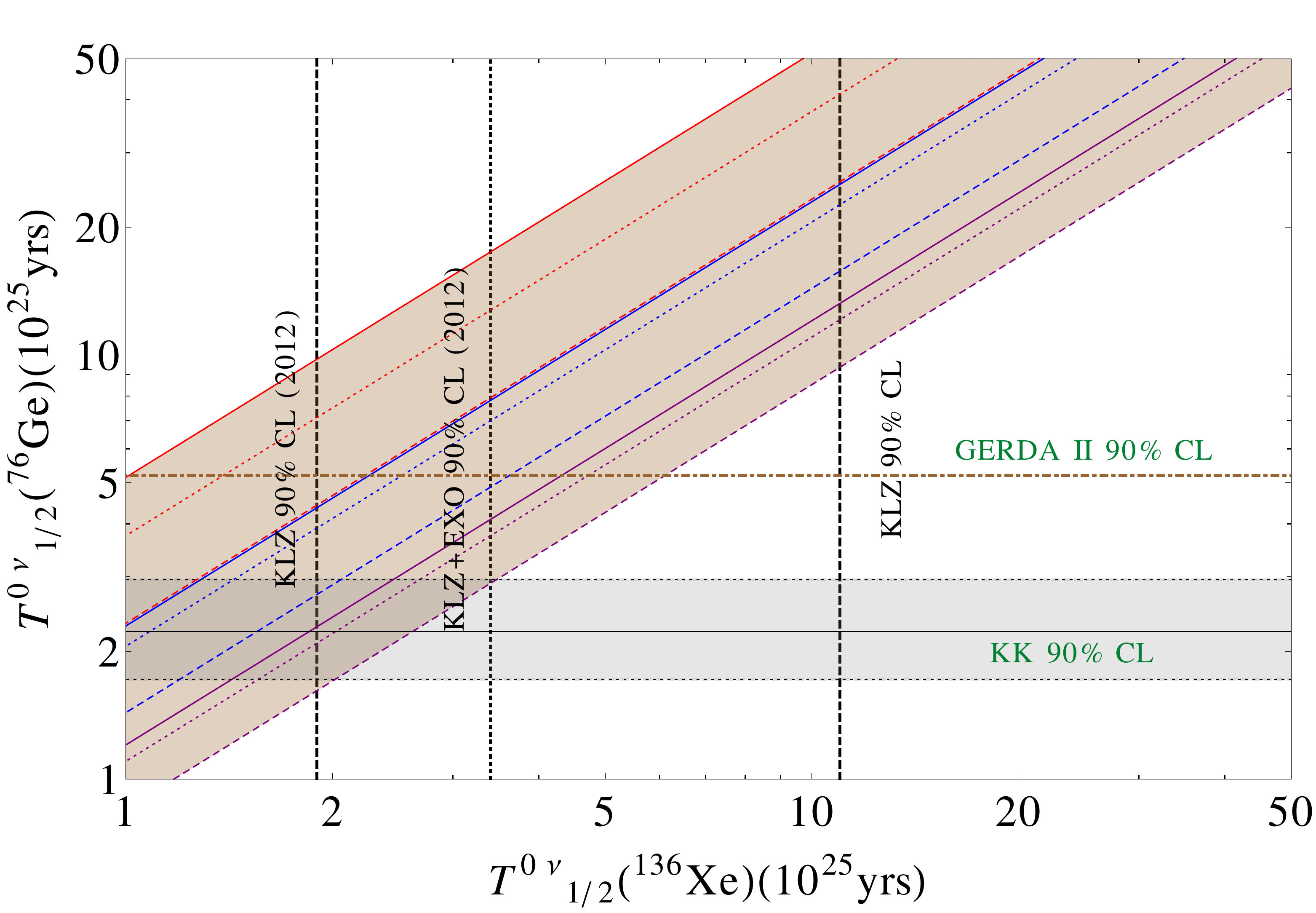}
\caption{The half-life $T^{0\nu}_{1/2}(^{76} \rm{Ge})$ vs  $T^{0\nu}_{1/2}(^{136} \rm{Xe})$. The brown shaded region represents the 
effect of NME uncertainty. The gray horizontal band represents the positive claim. The vertical lines represent the 
recent KLZ limit and the older KLZ limit, KLZ+EXO combined limit. The horizontal brown dot-dashed line represents the present limit from GERDA Phase-II \cite{gerda2}. }
\label{fig:correl}
\end{figure}

Till date, there has been only one claim of observation of $0\nu \beta \beta$ 
\cite{Klapdor:2006ff} for $^{76}\rm{Ge}$. The  limit from GERDA Phase-II rules out the  positive claim decisively. The validity of 
the positive claim has been  judged against the previous null result of KLZ and this has been shown that for all but one NME, the KLZ+EXO-200 combined limit 
rules out the positive claim \cite{Dev:2013vxa}. With the recent updated limit in hand, we re-check
the validity of the positive claim  against the null result of KLZ.  
 In Fig.~\ref{fig:correl}, we show the variation of the predicted half-life for $^{76}\rm{Ge}$ vs the half-life $^{136}\rm{Xe}$. The ratio of these 
two half-lives depend on the NME uncertainty and the phase space factor $G_{0\nu}$:
\be
\frac{T^{0\nu}_{1/2}(^{76}\rm{Ge})}{T^{0\nu}_{1/2}(^{136}\rm{Xe})}=\frac{G_{0\nu}(^{136}\rm{Xe}){\mathcal{M}}^2_{^{136}Xe}}{G_{0\nu}(^{76}\rm{Ge}){\mathcal{M}}^2_{^{76}Ge}}
\label{corel}
\ee
The different colored lines and the light purple region represent the predicted half-life for $^{76}\rm{Ge}$. The 
horizontal grey band represents the positive claim of observation of $0\nu \beta \beta$ in $^{76}\rm{Ge}$ \cite{Klapdor:2006ff}. The right most dashed 
vertical black line represent the most recent bound from KLZ. The other two vertical black lines represent the previous limits
from KamLAND-Zen and the combined limit from KamLAND-Zen+EXO-200 \cite{Gando:2012zm}. Note that, while the previous individual limit from KLZ
$T^{0\nu}_{1/2}=1.9 \times 10^{25}$ yrs didn't rule out the positive claim completely, the most recent improved limit decisively rules out the positive claim for all the NME compilations \cite{Rodriguez:2010mn, Menendez:2008jp, Barea:2013bz, Suhonen:2010zzc, Meroni:2012qf, 
Simkovic:2013qiy, Mustonen:2013zu}. 
\begin{figure*}[!ht]
\begin{center}
  \begin{minipage}[l]{2.0\columnwidth}
\includegraphics[width=17.0cm, height=5.25cm]{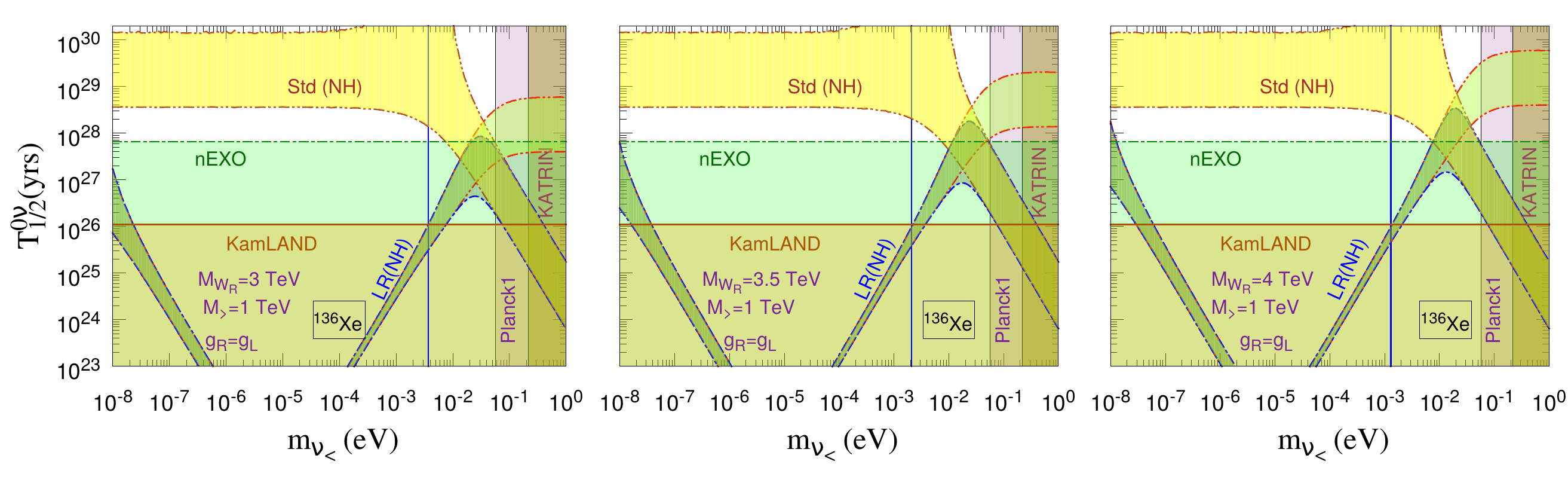}
\vspace*{-.5cm}
\includegraphics[width=17.0cm, height=5.25cm]{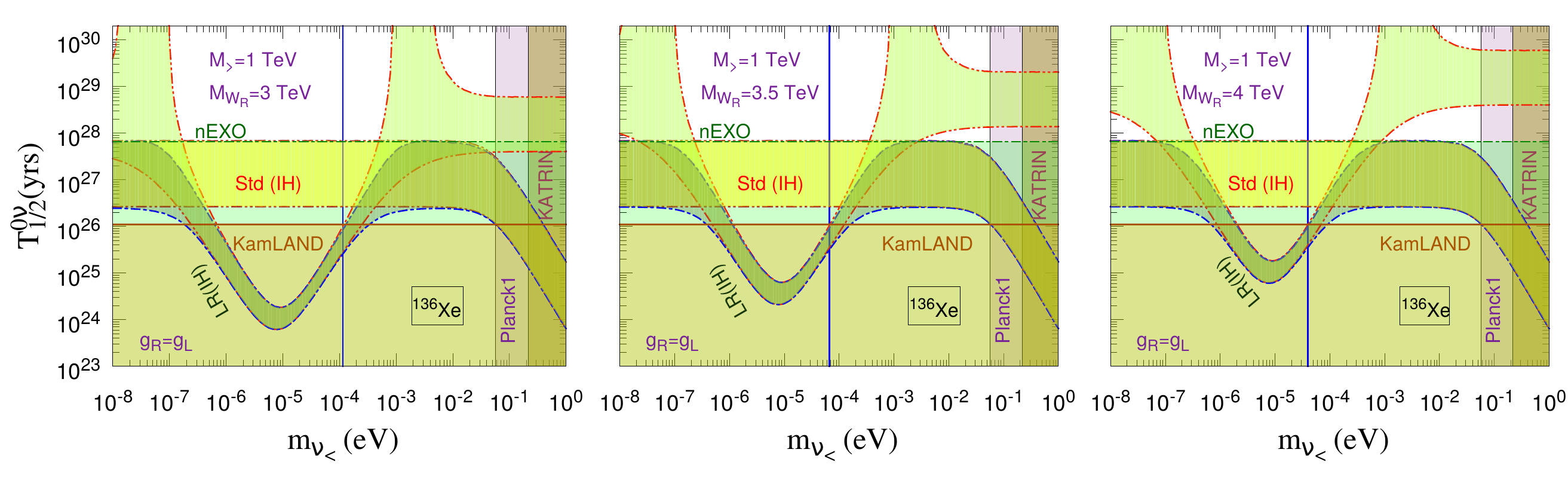}
\caption{The half-life $T^{0\nu}_{1/2}$ vs the light neutrino mass for NH (upper panel) and for IH (lower panel) for different 
$W_R$ masses. The area between
blue dot-dashed lines represents the total contributions (standard light neutrino+ heavy neutrino exchange) for Type-II dominance. }
\label{fig:hlife1a}
\end{minipage}
\end{center}
\end{figure*}
 
In  Fig.~\ref{fig:hlife1a}, we show the half-life $T^{0\nu}_{1/2}$ corresponding to the
light neutrino exchange by the yellow shaded region. The orange solid and the green dashed horizontal lines represent the 
KLZ limit $T^{0\nu}_{1/2}=1.07 \times 10^{26}$ yrs \cite{ KamLAND-Zen:2016pfg} and the projected sensitivity of future ton-scale experiment nEXO $T^{0\nu}_{1/2}=6.6 \times 10^{27}$\,yrs \cite{Albert:2014afa}. The vertical lines correspond to the stringent limit from PLANCK \cite{Ade:2015xua} and KATRIN \cite{Mertens:2015ila}. Note that, the new result  from KLZ  leaves only a small parameter space (the upper triangle at KLZ and KATRIN bounds crossing point) to be probed in  KATRIN. The nEXO can exhaust even more parameter space which is still allowed by PLANCK and can rule out the standard IH for all values of  light neutrino mass. 
\begin{table}[!h]
\begin{center}
\begin{tabular}{|l|l|l|l|}
\hline
\multicolumn{4}{|c|}{NH}  \\
\hline
\hline
\backslashbox{$g_R=g_L$}{$M_{W_R}$} &3 TeV & 3.5 TeV & 4 TeV \\
\hline
$m_{<}$ (eV) $\gtrsim$ & $3.7\times 10^{-3}$ & $2.1\times 10^{-3}$ & $1.3\times 10^{-3}$ \\ 
\hline
$M_{<}$ (GeV) $\gtrsim$ & 72.3 & 41.1 & 25.4 \\
\hline
$m_{<}$ (eV) $\lesssim$ & $2 \times 10^{-8}$ & $ 4 \times 10^{-8}$ & $8 \times 10^{-8}$  \\ 
\hline
$M_{<}$ (MeV) $\lesssim$ & $ 0.52$ & $ 0.93$ & $1.77$ \\
\hline
\hline
\multicolumn{4}{|c|}{IH}  \\
\hline
\hline
\backslashbox{$g_R=g_L$}{$M_{W_R}$} &3 TeV & 3.5 TeV & 4 TeV \\
\hline
$m_{<}$ (eV) $\gtrsim$ & $1.2\times 10^{-4}$ & $6.6\times 10^{-5}$ & $4\times 10^{-5}$ \\ 
\hline
$M_{<}$ (GeV) $\gtrsim$ & 2.25 & 1.3 & 0.78 \\
\hline
$m_{<}$ (eV) $\lesssim$ & $6.5 \times 10^{-7}$ & $1.2 \times 10^{-6}$ & $2 \times 10^{-6}$  \\ 
\hline
$M_{<}$ (MeV) $\lesssim$ & $12.8$ & $ 23.6$ & $39.3$ \\
\hline
\end{tabular}
\caption{The limits on the lightest neutrino mass $m_<$ and the lightest RH neutrino mass $M_<$, that come from KLZ, assuming a 
Type-II dominance in the lightest neutrino mass. 
}\label{tab:rhmassbounds}
\end{center}
\end{table}

In addition to the canonical light neutrino contribution, we also consider the RR contribution and show the prediction 
in the same figure, where we 
assume a Type-II dominance in the light neutrino mass matrix \cite{Tello:2010}.  For illustrative purpose, we consider a benchmark where the 
heaviest of the three RH neutrino $M_>=1$ TeV and the RH gauge boson has masses $M_{W_R}=3,\, 3.5$ and 4~TeV. We vary  the lightest light neutrino mass from $10^{-8}$ eV. The light green band covered by red dash-double-dot lines in the figure represents the RR exchange contribution. The area between dot-dashed blue lines that is shaded in deep green represents the total contribution that arises from the light neutrino exchange and the heavy neutrino-right handed gauge boson exchange. The KLZ result rules out significant amount of parameter space. For  NH scenario, 
the  lightest RH neutrino $M_{<}$ in between 0.52\,MeV-72.3\,GeV,\, 0.93\,MeV-41.1\,GeV,\, and 1.77\,MeV-25.4\,GeV are ruled out for $M_{W_R}=$3,\,3.5, and 4 TeV, respectively.  Similarly, for IH scenario, the ruled out mass 
ranges are  12.8\,MeV-2.25\,GeV, \,23.6\,MeV-1.3\,GeV, and 39.3\,MeV-0.78 \,GeV. A summary of these results is also given in Table-\ref{tab:rhmassbounds}. For 
 $g_R\neq g_L$, 
still assuming $f_L=f_R$ for simplicity, the result resembles  to Fig.~\ref{fig:hlife1a}. The summary of the results for this scenario of  $g_R=0.5 \neq g_L$ is given in   
Table-\ref{tab:gRis0pt5} (see the Appendix). 

{ For the $0\nu\beta\beta$ process mediated by the heavy RH neutrinos 
the energy scales at which effective 
interactions are generated and the scale at which the process are measured can be different. The QCD 
correction in the RG running and color mismatch due to these corrections may lead to 
substantial correction in $0\nu\beta\beta$ decays \cite{Mahajan:2013ixa, Peng:2015haa, Gonzalez:2015ady}. In Ref.~\cite{Gonzalez:2015ady},
 the authors have calculated the  leading 
QCD corrections to the complete set of short range $d=9$ $0\nu\beta\beta$ operators. 
For few of the operators the  corrections  can be large by order of magnitude. However, for 
 LR model, the corresponding operator  ($|C_3(\Lambda)|$) leads to small correction (see TABLE-II of \cite{Gonzalez:2015ady})
 to make any significant difference in our conclusions.}

\begin{table}[!h]
\begin{tabular}{|l|c|c|c|}
\hline
$|m^{\nu}_{ee}|$ and other BSM factors  & 
\multicolumn{3}{c|}{ Limits for $^{136}\rm{Xe}$ } \\ \hline
\multirow{2}{*}{Canonical: [eV]} & \multirow{2}{*}{Argonne}  & 
intm   &  $0.114$  \\ 
\cline{3-4}&  & large &  $0.095$\\ 
\cline{2-4}
\multirow{2}{*}{$|m^{\nu}_{ee}|=|\sum_i U_{ei}m_i|$}& \multirow{2}{*}{CD-Bonn}  & 
intm   &  $0.089$  \\ 
\cline{3-4}&  & large &  $0.078$ \\
\hline
\multirow{2}{*}{RR: [TeV$^{-5}$]}  & \multirow{2}{*}{Argonne}  & 
intm   &  $0.080$  \\ 
\cline{3-4}&  & large &  $0.082$\\ 
\cline{2-4}
\multirow{2}{*}{$\frac{1}{M^4_{W_R}}\left| \sum_i\frac{{V^*}^2_{ei} }{M_i} \right|$}& \multirow{2}{*}{CD-Bonn}  & 
intm   &  $0.078$  \\ 
\cline{3-4}&  & large &  $0.076$ \\
\hline
\multirow{2}{*}{LL: [TeV$^{-1}$]}  & \multirow{2}{*}{Argonne}  & 
intm   &  $3.34\times 10^{-6}$  \\ 
\cline{3-4}&  & large &  $3.42 \times 10^{-6}$\\ 
\cline{2-4}
\multirow{2}{*}{$\left| \sum_i\frac{{S^*}^2_{ei} }{M_i} \right|$}& \multirow{2}{*}{CD-Bonn}  & 
intm   &  $3.28\times 10^{-6}$  \\ 
\cline{3-4}&  & large &  $3.17\times 10^{-6}$ \\

\hline 
\multirow{2}{*}{Right triplet: [TeV$^{-5}$]} & \multirow{2}{*}{Argonne}  & 
intm   &  $4.54\times 10^{-4}$  \\ 
\cline{3-4}&  & large &  $4.65\times 10^{-4}$\\ 
\cline{2-4}
\multirow{2}{*}{$\frac{1}{{m^2_{\delta^{--}_R }M^4_{W_R} }}|\sum_i V^2_{ei}M_i| $}& \multirow{2}{*}{CD-Bonn}  & 
intm   &  $4.46\times 10^{-4}$  \\ 
\cline{3-4}&  & large &  $4.32\times 10^{-4}$ \\
\hline
\hline $\lambda$ exchange: [TeV$^{-2}$]& \multicolumn{3}{c|}{ $(3.18-4.05)\times 10^{-5}$} \\
$\frac{1}{M^2_{W_R}}\sum_i \left|U_{ei}T^*_{ei}  \right|$ & 
\multicolumn{3}{c|}{ } \\
\hline $\eta$ exchange: & \multicolumn{3}{c|}{ $(1.22- 1.38)\times 10^{-9}$  } \\
$\tan \xi \sum_i \left|U_{ei}T^*_{ei}  \right|$ & \multicolumn{3}{c|}{ } \\
\hline
\end{tabular}
\caption{The limits on the effective mass $m^{\nu}_{ee}$ and other relevant BSM parameters for LR symmetry that satisfy  KLZ \cite{KamLAND-Zen:2016pfg}. For all exchange mechanisms, other than $\lambda$ and $\eta$, we consider the NME from \cite{Meroni:2012qf}. For $\lambda$ and $\eta$, we follow \cite{Pantis:1996py, Barry:2013xxa}.  \label{tab:tab2}}
\end{table}

\begin{figure*}[!ht]
\begin{center}
\begin{minipage}[l]{2.0\columnwidth}
\includegraphics[width=17.0cm]{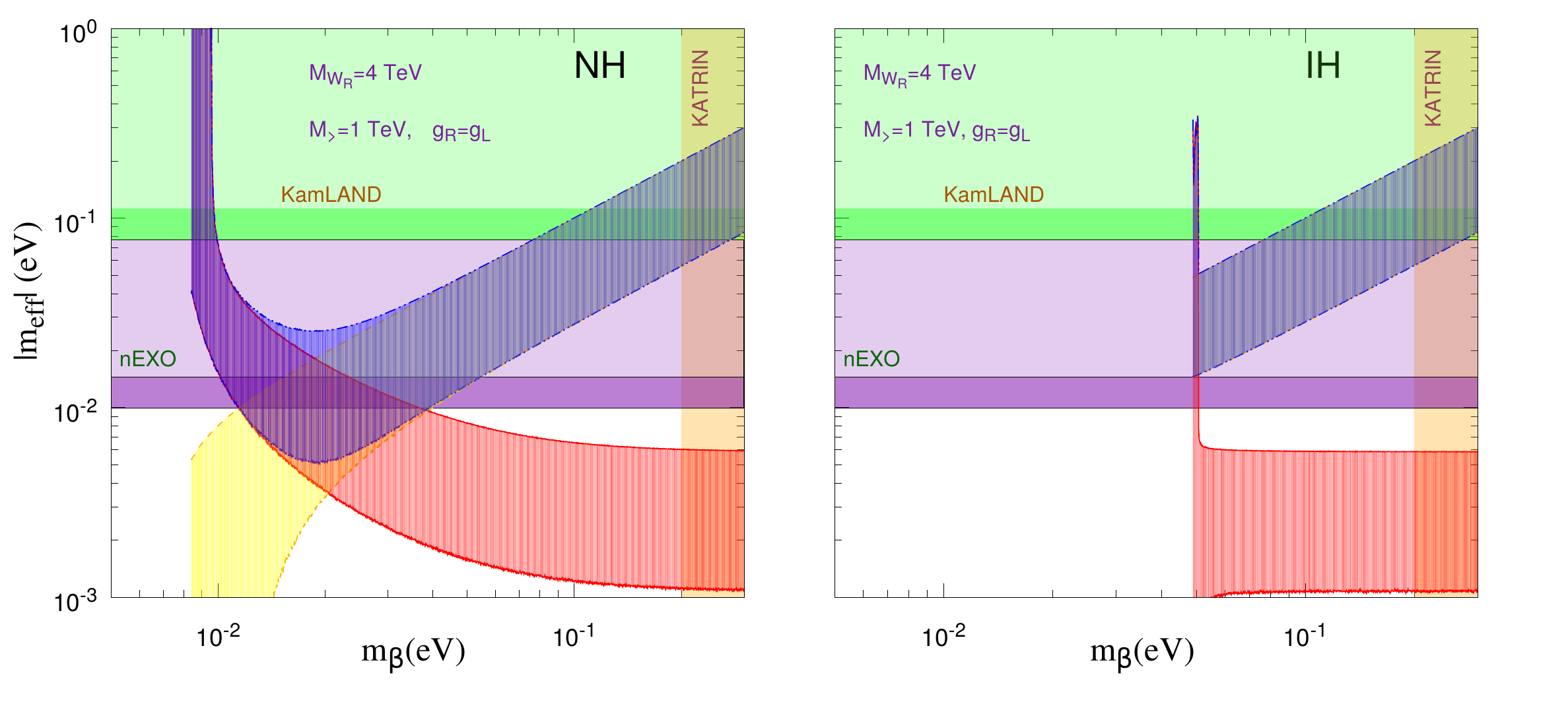}
\caption{The effective mass of $0\nu \beta \beta$ vs the effective mass of $\beta$-decay.The yellow and red region correspond to the
light neutrino exchange and RR contribution. The blue region represents the total contribution. The future ton-scale experiment nEXO can rule out the IH scenario for the adopted benchmark points. }
\label{fig:correl2}
\end{minipage}
\end{center}
\end{figure*} 
The future ton scale experiment nEXO will completely rule out any distinguishable NH BSM signature for the scenario $M_>=1\, \text{TeV}$ and $M_{W_R} \leq 4 $ TeV of Type-II dominance. In the IH scenario there is very limited scope to probe distinguishable Type-II dominant BSM physics in nEXO, that has  yet not been ruled out by KLZ. Both the  canonical or RR IH scenario can be completely ruled out by nEXO for even higher $W_R$ masses. Thus, may play a decisive role in fixing the mass  hierarchy. 
For the other contributions in LRSM that emerge from Type-I dominance, such as,  LL, $\eta
$, $\lambda$ or Higgs triplet contribution,  we show the  limits on the particle physics 
parameters  in Table.~\ref{tab:tab2} ( for $^{136}\rm{Xe}$ ) and for $^{76}\rm{Ge}$ in Table.~\ref{tab:tab2ge}.

While an individual measurement of $0\nu\beta\beta$ alone is not sufficient to 
conclusively point out the dominant mechanism behind this, the correlation 
of several other observables, such as observable for $\beta$ decay, the 
cosmological masses and the effective mass of $0\nu \beta \beta$ together 
might be 
indicative \cite{review}.  In Fig.~\ref{fig:correl2}, we show the correlation between the effective mass of light and heavy neutrino exchange in $0\nu \beta \beta$ and the observable of 
 $\beta$ decay. The yellow and pink regions represent the predictions for 
the light neutrino exchange and the RR exchange mechanisms in $0\nu\beta \beta$. Note that, for Type-II dominance and for 
mass of RH neutrino $M_>=1$ TeV, 
there is no other contributions in $\beta$ decay.  This is also evident from this figure, that the 
Type-II IH scenario can be completely ruled out by the ton scale experiments nEXO.

\begin{figure}[!b]
\begin{center}
\includegraphics[width=8.5cm]{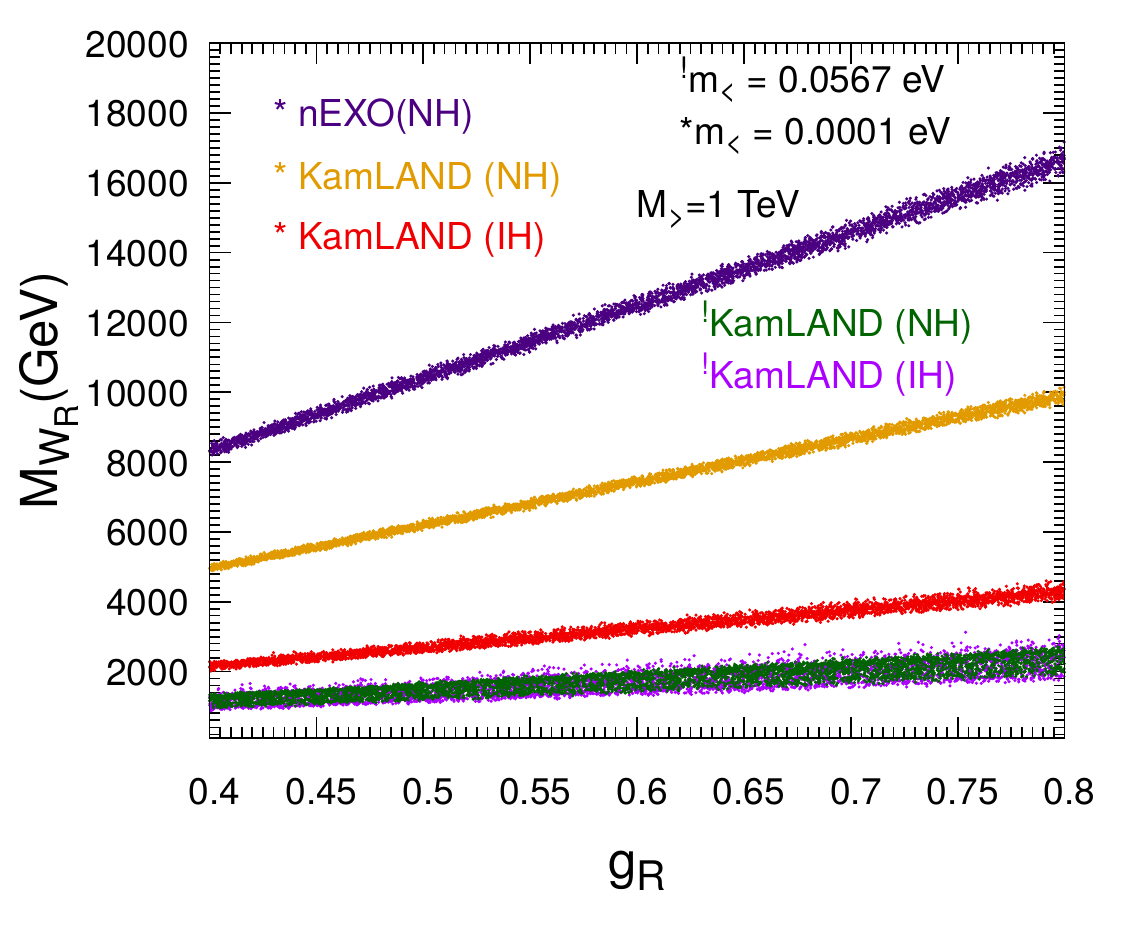}
\caption{The variation of $M_{W_R}$ vs the gauge coupling $g_R$, that satisfy the KLZ bound \cite{KamLAND-Zen:2016pfg} and the future sensitivity of nEXO \cite{Albert:2014afa}.}
\label{fig:grvsmwr}
\end{center}
\end{figure}

Note that, the weak gauge couplings of LR model may be different ($g_R\neq g_L$) if 
LR symmetry does not preserve the D-parity \cite{Chang:1983fu, Chang:1984uy}. An independent verification of $g_R \neq g_L$ in  $0\nu\beta\beta$  
is therefore  necessary to carry out. In Fig.~\ref{fig:grvsmwr}, we show the parameter space in $g_R-M_{W_R}$  
plane that saturates the KLZ limit and the projected sensitivity of nEXO, assuming the heavy 
neutrino mass $M_>=1$ TeV, and   $f_L=f_R$ for simplicity. In deriving the limits, we consider both the canonical contribution and RR together. We set the light neutrino masses $m_<$ near quasi degenerate 
regime 0.0567 eV that saturates the PLANCK limit $m_{\Sigma}=0.17$ eV \cite{Ade:2015xua}  and hierarchical regime $10^{-4}$ eV. The green (NH) and purple (IH) points 
represent the allowed-parameters that saturate KLZ limit for $m_<=0.0567$ eV. Note that
in quasi-degenerate regime where the dominant contribution comes from light neutrino exchange, the two scenarios overlap, making this difficult to determine the hierarchy.  The red (IH) and orange (NH) points 
represent the allowed-parameters that saturate the KLZ limit for much hierarchical mass 
regime $m_<=10^{-4}$ eV, that is clearly separable. The blue dots correspond to  optimal nEXO sensitivity.
Note that, for the quasi-degenerate  light neutrino mass, the total contribution 
from canonical and RR exchange can  be ruled out for much less sensitivity of nEXO, and  hence  does 
not show up in the figure where  we have considered $T^{0\nu}_{1/2}=6.6 \times 10^{26}$ yrs. The same happens for IH hierarchical  scenario.
For each of the selected light neutrino mass, 
the band in the $g_R-M_{W_R}$ plane corresponds to the variation of the oscillation parameters in their 3$\sigma$ region \cite{Gonzalez-Garcia:2015qrr}.
In addition, we also include the NME variation \cite{Meroni:2012qf}. \\

{Note that, although we have explored the limits from $ 0\nu \beta \beta$, there are other relevant seahces, namely, the dijet searches at LHC for $W^{\prime}$ \cite{ATLAS:2016lvi}, the same sign dilepton searches \cite{Aad:2015xaa, Khachatryan:2014dka} give stringent constraints on the masses of the   gauge boson $W_R$ and heavy neutrino $N$ and. For a  summary of relevant searches, see  \cite{Helo:2015ffa}. The updated 13 TeV dijet search from ATLAS (assuming $75\%$ branching ratio for $W'\to j j$ and an SM like coupling )  rules out the $W_R$ mass upto 2.9 TeV  \cite{ATLAS:2016lvi}. For $M_N < M_{W_R}$, this search has a very minor dependency on the mass of $N$ ( through branching ratio ). Assuming that the $W_R$ couples to  two light generation of quarks through CKM type mixing, the branching ratios to different states become:  ${\rm{Br}}(W \to e N)=10\%$,  ${\rm{Br}}(W_R \to j j) $ and ${\rm{Br}}(W_R \to t b)$  as  $60 \%$ and 30$\%$, respectively \cite{Mitra:2016kov}. For $60 \%$ branching ratio, the limit is similar $M_{W_R} \sim 2.8$ TeV. On the other hand, the search from same-sign dilepton constrains
the TeV-hundred GeV $M_{W_R}-M_N$ mass plane. The $95 \%$ C.L limit from ATLAS 8 TeV search on the $M_{W_R}$  reaches 2.9 TeV for heavy neutrino mass $M_N =50$ GeV \cite{Aad:2015xaa} (see Fig.~11 and Table.~6 of \cite{Aad:2015xaa} for  a scan over parameter space).  For all the mediators $M_{W_R}$ and $M_N$ in the TeV-few hundred GeV mass range, the LHC same sign lepton search is most constraining and even gives much more stringent limits than $0 \nu \beta \beta$ \cite{Deppisch:2012nb}. However, for lower $N$ masses, such as only few GeV, or MeV and the other mass range where $M_N > M_{W_R}$, the LHC same sign dilepton search is not applicable \cite{Dev:2013vxa}, therefore allowing a huge range of parameters, where $0\nu \beta \beta$ can be more informative.   }

Below, we discuss the other BSM scenario RPV susy. \\

{\bf{\textit{R-Parity Violation:} }}
The RPV superpotential with the $\lambda'$ coupling  is 
\begin{equation}
  W_{\not{R}}= \lambda' LQd^c. 
  \label{eqrpv}
\end{equation}
This induces the following Lagrangian terms,
\begin{equation}
  \mathcal{L}=-\lambda'_{ijk} \tilde{l}_i u_j d^c_k-\lambda'_{ijk} \tilde{u}_j l_i d^c_k+
\lambda'_{ijk} \tilde{d}_j \nu_{l_i} d^c_k+ \tilde{\nu}_{l_i} d_j d^c_k+ ... \label{eq:L}
\end{equation}

The $0 \nu \beta \beta$ process receives contribution from neutrino-sbottom exchange - via $\lambda'_{113}$, $\lambda'_{131}$ couplings, 
squark-gluino exchanges - via $\lambda'_{111}$ coupling \cite{msnu2beta, mwex, pionex, hirsch, allanach, vogel, Dev:2016uxj}. 
Here, we update the bounds on the relevant dimensionless parameters $\eta_{\tilde{q}}$ (relevant for sbottom exchange) and 
$\eta_{\tilde{g}}$ following the KLZ result. 
\begin{itemize}

\item
neutrino-squark contribution depends on the squark masses and the product of couplings $\lambda_{113} \lambda_{131}$. The dimensionless parameter for sbottom exchange is, 
$\eta_{\tilde{b}}= \frac{\lambda'_{113}\lambda'_{131}}{2 \sqrt{2} G_F}\sin 2\theta \left(\frac{1}{m^2_{\widetilde{b}_{1}} }-\frac{1}{m^2_{\widetilde{b}_{2}} }\right)$, where $\theta$ is the mixing between left and right handed chiral sbottom states $\tilde{b}_L$ and $\tilde{b}_R$. The 
masses of the physical sbottom states are $m_{\tilde{b}_1}$ and $m_{\tilde{b}_2}$, respectively. Similar to the sbottom exchange, other squarks
can also contribute in this process. 

\item
The gluino exchange can give large contribution in $0\nu \beta \beta$. Assuming, the gluino and the squarks as the mediators, 
the relevant dimensionless parameter $\eta_{\tilde{g}}$ is 
$\eta_{\tilde{g}}=\frac{\pi \alpha_s}{6} \frac{{\lambda^{\prime}}^2_{111}}{G^2_F m^4_{\tilde{d}_R}} \frac{m_p}{m_{\tilde{g}}}$, where for simplicity
we have assumed down-type squark exchange gives large contribution. Similar contribution can be obtained from  up-type squark exchange. 

\begin{figure}[!h]
\begin{center}
\includegraphics[width=8.400cm, height=6cm]{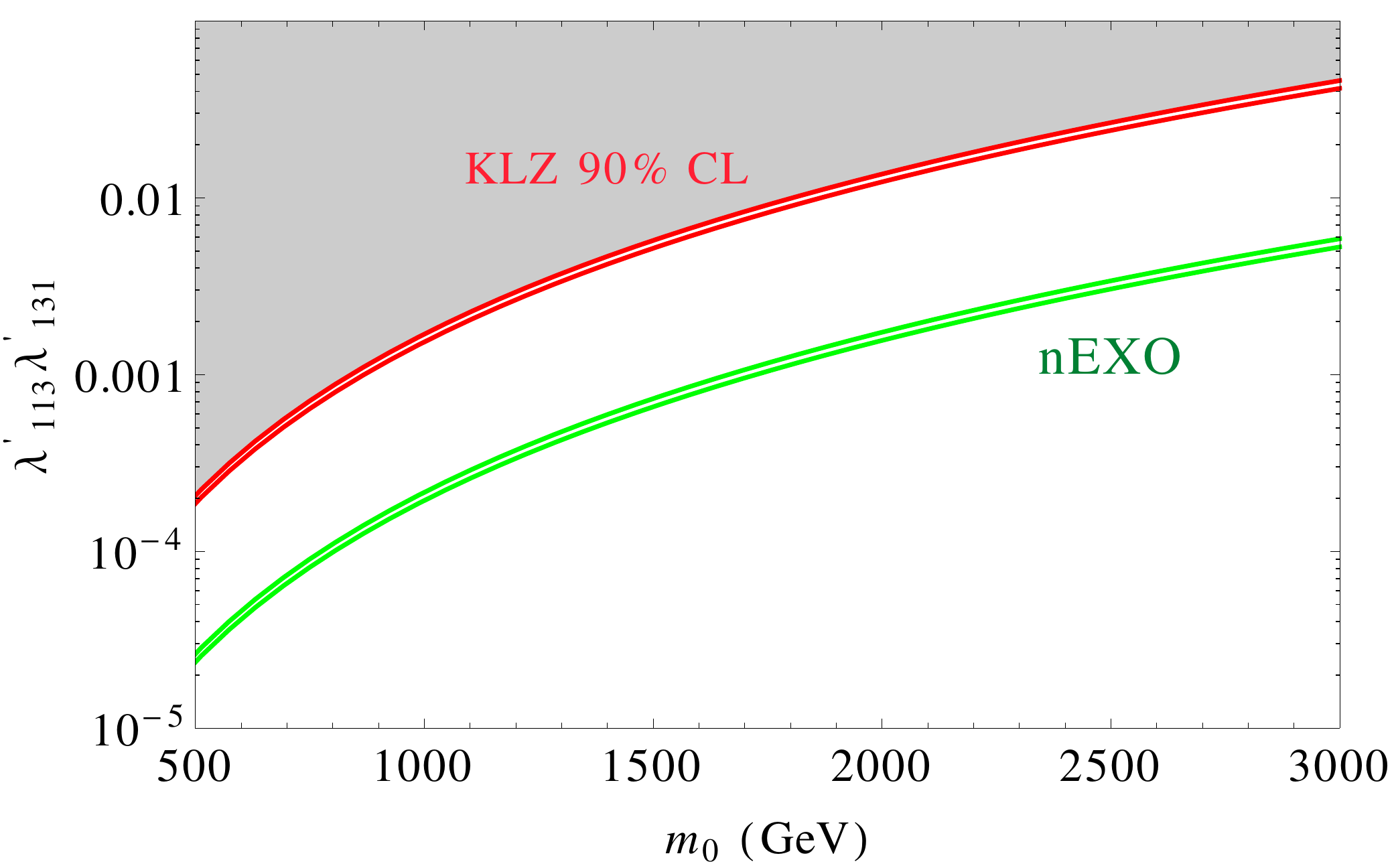}
\caption{The variation of $\lambda'_{113}\lambda'_{131}$ vs the common sbottom mass $m_0$ that satisfy the KLZ bound \cite{KamLAND-Zen:2016pfg} and saturate the future sensitivity of nEXO \cite{Albert:2014afa}. The different of the sbottom mass from $m_0$ is 60 GeV and the mixing $\sin 2\theta=10^{-4}$.}
\label{fig:rpvsmwr}
\end{center}
\end{figure} 
\begin{figure}[h!]
\begin{center}
\includegraphics[width=8.4000cm, height=6cm]{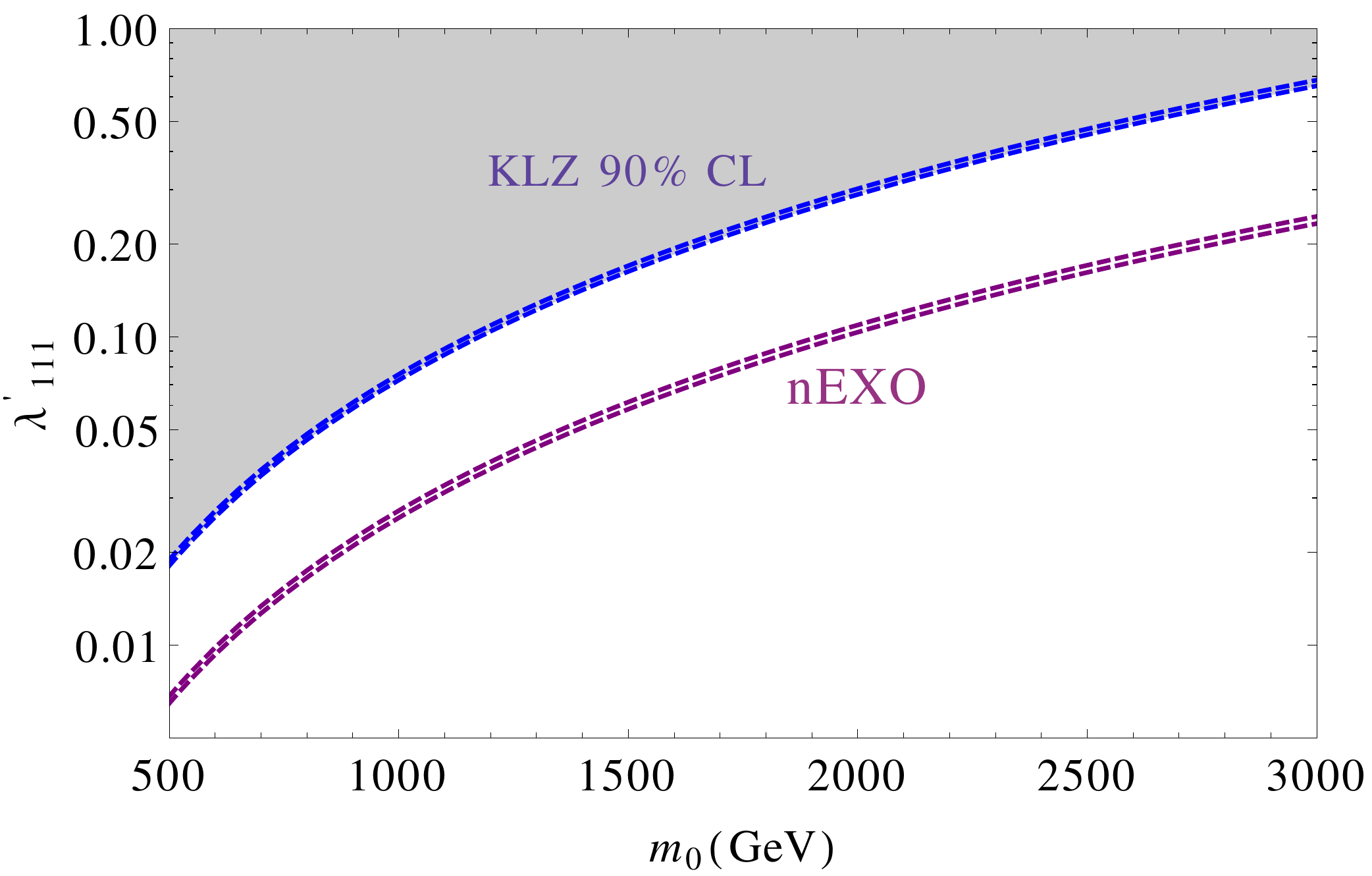}
\caption{The variation of $\lambda'_{111}$ vs the down type squark  mass $m_0$ that satisfy the KLZ bound \cite{KamLAND-Zen:2016pfg} and saturate the future sensitivity of nEXO \cite{Albert:2014afa}. The gluino mass has been set to 2.0 TeV.}
\label{fig:rpvglu}
\end{center}
\end{figure} 
\end{itemize}

In 
Table.~\ref{tab:tab3}, we provide the updated limits on the dimensionless  parameters $\eta_{\tilde{b}}$ and $\eta_{\tilde{g}}$. The limits corresonding to GERDA Phase-II measurements are given in Table.~\ref{tab:tab3ge}.
In Fig.~\ref{fig:rpvsmwr}, we show the constraints on $\lambda_{113} \lambda_{131}$ with respect to sbottom mass variation, that 
satisfies KLZ limits and saturates the nEXO sensitivity.  We consider the two sbottom masses
${m}_{\tilde{d}_1, \tilde{d}_2}=m_0\pm\Delta m$, with the mass difference being $\Delta m=60$ GeV and  the mixing 
$\sin 2 \theta=10^{-4}$. The gray shaded region is ruled out
by the KLZ limit.  In Fig.~\ref{fig:rpvglu} we show the limit on the $\lambda'_{111}$ that
corresponds to the gluino-down type squark exchange. The green/purple band represent the sensitivity of the future ton-scale experiment nEXO that can probe 
much lower regime of couplings 
$\lambda_{113} \lambda_{131}$/$\lambda'_{111}$. {In the analysis on RPV, we have not considered the effect of RG running. The 
gluino and squark exchange diagram will lead to operator mixing (tensorial and pseudoscalar). The QCD corrections to the coefficients 
of pseudo-scalar effective operator are large while for tensorial operators, the correction is not significant as shown in \cite{Gonzalez:2015ady}. Complete analysis including mixing between pseudoscalar and tensor operator would be more involved and will be considered in future work.}  \\

\begin{table}[!h]
\centering
\begin{tabular}{|c|c|c|c|}
\hline
RPV  & 
\multicolumn{3}{c|}{ Limits for $^{136}$Xe } \\ \hline
\hline
\multirow{4}{*}{$\eta_{\tilde{b}}$} & \multirow{2}{*}{Argonne}  & 
intm   &  $1.23\times 10^{-9}$  \\ 
\cline{3-4}&  & large &  $1.23\times10^{-9}$\\ 
\cline{2-4}
& \multirow{2}{*}{CD-Bonn}  & 
intm   &  $1.24\times 10^{-9}$  \\ 
\cline{3-4}&  & large &  $1.12\times10^{-9}$ \\
\hline
\hline
\multirow{4}{*}{$\eta_{\tilde{g}}$} & \multirow{2}{*}{Argonne}  & 
intm   &  $1.20\times 10^{-9}$  \\ 
\cline{3-4}&  & large &  $1.20\times10^{-9}$\\ 
\cline{2-4}
& \multirow{2}{*}{CD-Bonn}  & 
intm   &  $1.21\times 10^{-9}$  \\ 
\cline{3-4}&  & large &  $1.11\times10^{-9}$ \\
\hline
\end{tabular}
\caption{The limits on the RPV dimensionless parameters from KLZ \cite{KamLAND-Zen:2016pfg}.  We adopt the NME from \cite{Meroni:2012qf}. \label{tab:tab3}}
\end{table}

{Before conclusion, we would like to make some remarks about the searches for RPV at colliders. Introducing non-zero RPV coupling generally weakens up the mass and cross-section limits for the sparticles. The summary of the searches can be found in \cite{ATLAS:2015gma}, where the main focus is on pair production of squarks and gluinos, and their further decays. Several searches have been conducted for RPV $\lambda^{\prime \prime}$, $\lambda^{\prime}$ and $\lambda$ couplings. Limits have been set on the gluino mass $m_{\tilde{g}} \ge 1550$ GeV, where gluino decays to  
hadronic final states \cite{ATLAS:2016nij} via  via neutralino $\tilde{\chi}^0$ and $\lambda^{\prime \prime}$ coupling. On the other hand, for $\lambda$ coupling, searches have been conducted for fully leptonic channel 
\cite{ATLAS:2016soo}, originating from chargino decays. In the most constraining scenario, chargino mass upto 1.14 TeV have been excluded. The limit weakens for large mass hierarchy between chargino and lightest neutralino mass, where the decay products are boosted \cite{ATLAS:2016soo}. For a summary of Run-1 search see \cite{ATLAS:2015gma}. For $\lambda^{\prime}$ searches, a) searches for the third generation of squarks $p p \to \tilde{t} \tilde{t}^* \to bl bl$ constrain stop mass 1 TeV, where $\tilde{t}$ decays to b quark and $e$ \cite{ATLAS:2015jla}, b) searches for multilepton and b-jet constrain $\tilde{t}$ mass 1 TeV \cite{Chatrchyan:2013xsw}( relevant for $\lambda^{\prime}_{233}$ and $\lambda^{\prime}_{231}$ ). c) final states with $\tau$ and b-jets ( relevant  for $\lambda^{\prime}_{333}$ and $\lambda^{\prime}_{3jk}$) \cite{Khachatryan:2014ura}. For this search, assuming a 100 $\%$ branching ratio, mass limits have been set on stop mass as 580 GeV. Note that none of these above searches distinguishably  constrain $\lambda^{\prime}_{111}$ or $\lambda^{\prime}_{131}$ coupling. However, if more than one  RPV coupling is present, then the above mentioned searches will be relevant. In \cite{ATLAS:2015gma}, the channel that has been analysed is $ p p \to \tilde{g}\tilde{g} \to \tilde{\chi}^0_1 q q \tilde{\chi}^0_1  q q $ and $ p p \to \tilde{q}\tilde{q} \to \tilde{\chi}^0_1 q q \tilde{\chi}^0_1 $ with $\tilde{\chi}^0_1 \to l/\nu q q$. In most of the cases, this set limit on squark/gluino mass $m_{\tilde{g}}/m_{\tilde{q}} > 1$  TeV. We consider the gluino mass $\tilde{g}$ as 2 TeV in our analysis. The parameter space shown in Fig.~\ref{fig:rpvsmwr} and Fig.~\ref{fig:rpvglu} is consistent with the present LHC searches.   }

{\textit{\textbf{Conclusion}}}-- 
In the light of the recent results from KamLAND-Zen and GERDA Phase-II, we  update limits on the effective 
mass and new physics parameters for  $0\nu\beta\beta$ considering two widely discussed 
 BSM scenarios, Left-Right symmetry and RPV susy. 
The recent KLZ
limit puts stringent constraint on the effective mass for light neutrino exchange $m^{\nu}_{ee} \le 0.06-0.14$ eV, almost factor of two 
improvement as compared to the previous limit. We re-check the validity of 
the positive claim against the null result of  KLZ and
find that assuming  the light neutrino exchange is the only mechanism of $0\nu\beta \beta$, the recent KLZ and GERDA Phase-II  limits individually rule out the positive claim of observation completely, for all NME. In the BSM scenarios, our findings are i) 
the KLZ limit provides factor of two  improvement for the BSM parameters ii)
For Type-II dominated 
Left-Right model,  a wide range of RH neutrino mass is now excluded. We show this explicitly for the RH gauge boson  mass $\le $ 4 TeV and the heaviest RH neutrino mass 1 TeV. This leaves  very limited parameter space to probe  distinguishable BSM contribution that comes from RR exchange. iii) For the RPV susy, couplings of order 
$\mathcal{O}(0.01-0.1)$ is ruled out from KLZ for sbottom/squark mass between 500 GeV-3 TeV iv)   We  show that the next generation ton scale 
experiment nEXO will be able to rule out distinguishable BSM signature for 
$M_{W_R}\leq 4$\,TeV in NH scenario and will be able to completely rule out IH scenario for BSM as well as Standard $0\nu\beta\beta$ contribution. For RPV scenarios, nEXO can probe $< \mathcal{O}( 10^{-2}$) couplings.

\section{Appendix \label{app1}}
The bounds on RH neutrino masses corresponding to Fig.~\ref{fig:hlife1a} and 
Table~\ref{tab:rhmassbounds} but, for $g_R=0.5$ are summarized in this table. 
\begin{table}[!htb]
\begin{center}
\begin{tabular}{|l|l|l|l|}
\hline
\multicolumn{4}{|c|}{NH}  \\
\hline
\hline
\backslashbox{$g_R=0.5$}{$M_{W_R}$}
& 2.5 TeV & 3.0 TeV & 3.5 TeV \\
\hline
$m_{<}$ (eV) $\gtrsim$ & $2.8\times 10^{-3}$ & $1.4\times 10^{-3}$ & $8\times 10^{-4}$ \\ 
\hline
$M_{<}$ (GeV) $\gtrsim$ & 54.8 & 27.4 & 15.7 \\
\hline
$m_{<}$ (eV) $\lesssim$ & $3 \times 10^{-8}$ & $6 \times 10^{-8}$ & $10^{-7}$  \\ 
\hline
$M_{<}$ (MeV) $\lesssim$ & $0.6$ & $1.2$ & $2.1$ \\
\hline
\hline
\multicolumn{4}{|c|}{IH}  \\
\hline
\hline
\backslashbox{$g_R=0.5$}{$M_{W_R}$} 
&2.5 TeV & 3.0  TeV & 3.5 TeV \\
\hline
$m_{<}$ (eV) $\gtrsim$& $9\times 10^{-5}$ & $4\times 10^{-5}$ & $2\times 10^{-5}$ \\ 
\hline
$M_{<}$ (GeV) $\gtrsim$ & 1.8 & 0.8 & 0.4 \\
\hline
$m_{<}$ (eV)  $\lesssim$ & $9 \times 10^{-7}$ & $2 \times 10^{-6}$ & $4\times 10^{-6}$  \\ 
\hline
$M_{<}$ (MeV) $\lesssim$ & $ 18.7$ & $ 41.6$ & $ 83.3$ \\
\hline
\end{tabular}
\caption{The bounds on the RH and lightest neutrino masses from KLZ \cite{KamLAND-Zen:2016pfg}. In NH case lightest right handed neutrino $M_{<}$ is either lighter 
than 0.6/1.2/2.1 MeV or heavier than 
54.8/27.4/15.7 GeV for $M_{W_R}=$2.5/3/3.5 TeV and $M_{>}=1$ TeV. Similarly,  In IH case $M_{<}$ is either lighter than 18.7/41.6/83.3 MeV or heavier than 1.8/0.8/0.4 GeV for $M_{W_R}=$2.5/3/3.5 TeV and $M_{>}=1$ TeV.}\label{tab:gRis0pt5}
\end{center}
\end{table}

\begin{table}[!h]
\begin{tabular}{|l|c|c|c|}
\hline
$|m^{\nu}_{ee}|$ and other BSM factors  & 
\multicolumn{3}{c|}{ Limits for $^{76}\rm{Ge}$ } \\ \hline
\multirow{2}{*}{Canonical: [eV]} & \multirow{2}{*}{Argonne}  & 
intm   &  $0.196$  \\ 
\cline{3-4}&  & large &  $0.172$\\ 
\cline{2-4}
\multirow{2}{*}{$|m^{\nu}_{ee}|=|\sum_i U_{ei}m_i|$}& \multirow{2}{*}{CD-Bonn}  & 
intm   &  $0.183$  \\ 
\cline{3-4}&  & large &  $0.160$ \\
\hline
\multirow{2}{*}{RR: [TeV$^{-5}$]}  & \multirow{2}{*}{Argonne}  & 
intm   &  $0.200$  \\ 
\cline{3-4}&  & large &  $0.176$\\ 
\cline{2-4}
\multirow{2}{*}{$\frac{1}{M^4_{W_R}}\left| \sum_i\frac{{V^*}^2_{ei} }{M_i} \right|$}& \multirow{2}{*}{CD-Bonn}  & 
intm   &  $0.133$  \\ 
\cline{3-4}&  & large &  $0.113$ \\
\hline
\multirow{2}{*}{LL: [TeV$^{-1}$]}  & \multirow{2}{*}{Argonne}  & 
intm   &  $8.36\times 10^{-6}$  \\ 
\cline{3-4}&  & large &  $7.35 \times 10^{-6}$\\ 
\cline{2-4}
\multirow{2}{*}{$\left| \sum_i\frac{{S^*}^2_{ei} }{M_i} \right|$}& \multirow{2}{*}{CD-Bonn}  & 
intm   &  $5.54\times 10^{-6}$  \\ 
\cline{3-4}&  & large &  $4.73\times 10^{-6}$ \\

\hline 
\multirow{2}{*}{Right triplet: [TeV$^{-5}$]} & \multirow{2}{*}{Argonne}  & 
intm   &  $4.58\times 10^{-4}$  \\ 
\cline{3-4}&  & large &  $9.99\times 10^{-4}$\\ 
\cline{2-4}
\multirow{2}{*}{$\frac{1}{{m^2_{\delta^{--}_R }M^4_{W_R} }}|\sum_i V^2_{ei}M_i| $}& \multirow{2}{*}{CD-Bonn}  & 
intm   &  $7.54\times 10^{-4}$  \\ 
\cline{3-4}&  & large &  $6.43\times 10^{-4}$ \\
\hline
\hline $\lambda$ exchange: [TeV$^{-2}$]& \multicolumn{3}{c|}{ $(1.61-7.51)\times 10^{-5}$} \\
$\frac{1}{M^2_{W_R}}\sum_i \left|U_{ei}T^*_{ei}  \right|$ & 
\multicolumn{3}{c|}{ } \\
\hline $\eta$ exchange: & \multicolumn{3}{c|}{ $(2.87- 7.77)\times 10^{-9}$  } \\
$\tan \xi \sum_i \left|U_{ei}T^*_{ei}  \right|$ & \multicolumn{3}{c|}{ } \\
\hline
\end{tabular}
\caption{The limits on the effective mass $m^{\nu}_{ee}$ and other relevant BSM parameters for LR symmetry that satisfy  GERDA-II  \cite{gerda2}. For all exchange mechanisms, other than $\lambda$ and $\eta$, we consider the NME from \cite{Meroni:2012qf}. For $\lambda$ and $\eta$, we follow \cite{Pantis:1996py, Barry:2013xxa}.  \label{tab:tab2ge}}
\end{table}

\begin{table}[!h]
\centering
\begin{tabular}{|c|c|c|c|}
\hline
RPV  & 
\multicolumn{3}{c|}{ Limits for $^{76}\rm{Ge}$ } \\ \hline
\hline
\multirow{4}{*}{$\eta_{\tilde{g}}$} & \multirow{2}{*}{Argonne}  & 
intm   &  $3.07\times 10^{-9}$  \\ 
\cline{3-4}&  & large &  $2.54\times10^{-9}$\\ 
\cline{2-4}
& \multirow{2}{*}{CD-Bonn}  & 
intm   &  $2.98\times 10^{-9}$  \\ 
\cline{3-4}&  & large &  $2.51\times10^{-9}$ \\
\hline
\multirow{4}{*}{$\eta_{\tilde{g}}$} & \multirow{2}{*}{Argonne}  & 
intm   &  $3.11\times 10^{-9}$  \\ 
\cline{3-4}&  & large &  $2.61\times10^{-9}$\\ 
\cline{2-4}
& \multirow{2}{*}{CD-Bonn}  & 
intm   &  $3.55\times 10^{-9}$  \\ 
\cline{3-4}&  & large &  $3.07\times10^{-9}$ \\
\hline
\end{tabular}
\caption{The limits on the RPV dimensionless parameters from GERDA Phase-II \cite{gerda2}.  We adopt the NME from \cite{Meroni:2012qf}. \label{tab:tab3ge}}
\end{table}

\acknowledgements
{\textit{\textbf{Acknowledgments}}}-- 
M.M acknowledges the hospitality of IPPP, Durham University, UK, where this work has been completed. The work of M.M is partially supported
by the DST INSPIRE Grant INSPIRE-15-0074.

\end{document}